\documentclass[12pt,preprint]{aastex}
\usepackage{graphics}

\begin{document}

\title{Galactic Cosmic Rays from Superbubbles and the Abundances of Lithium, Beryllium, and Boron}

\author{Andreu Alib\'es, Javier Labay, Ramon Canal\altaffilmark{1}}

\affil{Departament d'Astronomia i Meteorologia, Universitat de Barcelona,
           {Mart\'\i} i Franqu\`es 1, 08028 Barcelona, Spain}
\altaffiltext{1}{Institut d'Estudis Espacials de Catalunya, Edifici Nexus, Gran
	   Capit\`a 2--4, 08034 Barcelona, Spain}
\email{email: aalibes, javier, ramon@am.ub.es}

\shorttitle{GCR from SB and LiBeB evolution}

\begin{abstract} In this article we study the galactic evolution of the LiBeB
elements within the framework of a detailed model of the chemical evolution of the
Galaxy that includes galactic cosmic ray nucleosynthesis by particles accelerated in
superbubbles. The chemical composition of the superbubble consists of varying
proportions of ISM and freshly supernova synthesized material. The observational
trends of $^{6} $LiBeB evolution are nicely reproduced by models in which GCR come
from a mixture of 25\% of supernova material with 75\% of ISM, except for $^{6}
$Li,  for which maybe an extra source is required at low metallicities. To account
for  $^{7} $Li evolution several additional sources have been considered
(neutrino--induced nucleosynthesis, nova outbursts, C--stars). The model fulfills
the energetic requirements for GCR acceleration. 
\keywords{cosmic rays --- nuclear reactions, nucleosynthesis, abundances --- stars:
abundances}  
\end{abstract} 
\maketitle

\section{INTRODUCTION}  The evolution of the light nuclides ($ ^{6,7} $Li, $ ^{9}
$Be, and $ ^{10,11} $B) has been, since the early 70's, a major topic in the studies
of galactic chemical evolution. The theoretical scenario for the synthesis of
lithium, beryllium, and boron has remained unchanged for years since the pioneering
work of Meneguzzi, Audouze, \& Reeves (1971). Recent observations (see Molaro et al.
1997; Boesgaard et al. 1999b, for instance) have revealed an almost linear
correlation of both Be and B with Fe at low metallicities. This new trend seems to
rule out the common view of galactic cosmic ray (GCR) particles being  accelerated
from material of the interstellar medium (ISM), as the latter predicts a quadratic
dependence on metallicity, and it has motivated models of LiBeB nucleosynthesis that
consider new sites for the production and acceleration of the GCR (Ramaty et al.
2000; Fields et al. 2000; Parizot \& Drury 2000). A source of low--energy cosmic
rays, enriched in C and O, was one of the first suggested solutions to the problem,
mainly inspired by gamma--ray observations in Orion which, however, once revised
appeared spurious (Bloemen et al. 1999). That has led, temporarily at least, to
discard the low--energy component hypothesis. 

Scenarios where the composition of the GCR was identical to that of the ejecta from
individual supernovae have also been ruled out by observations with the \emph{Cosmic
Ray Isotope Spectrometer} (CRIS) on board of the \emph{Advanced Composition
Explorer} (ACE) that measured the abundance of the radioactive isotope $ ^{59} $Ni
and its decay product $ ^{59} $Co in current epoch GCR (Binns et al. 1999). $ ^{59}
$Ni decays by electron capture with a half life of 7.6$\times$10$ ^{4} $ years, but
it does not decay if it is accelerated. The abundance measurements show a very small
amount of $ ^{59} $Ni in GCR, and so there must be a minimum delay of 10$ ^{5} $
years between the explosive nucleosynthesis of the material entering in GCR and its
acceleration. Such a delay, however, could be accounted for in models where the GCR
originated in a superbubble (SB) region of hot, rarefied, metal--rich gas swept out
by the collective effects of massive star winds and supernova explosions.

In the SB scenario of Ramaty et al. (2000)  LiBeB nucleosynthesis is a primary
mechanism (i.e., production $\propto$ O), as the composition of the GCR, that
exclusively comes  from the ejecta of  supernovae inside the SB, does not change in
time, and it is quite metal--rich. Other authors preferred a combination of a
primary and a secondary (i.e., production $\propto$ O$ ^{2} $) mechanism.  Parizot
\& Drury (2000) considered also a SB scenario for the acceleration of GCR, but they
adopted a composition of the SB material that is a mixture of ISM plus a 2 or 3\% of
SN ejecta, according to non--magnetic SB models.  On the other hand, Fields et al.
(2000) combined two sources of GCR: one accelerated from the ISM, that has a
secondary character, and a second component coming from SB and made of pure ejecta
of SNe, having thus a primary behaviour. In the last two cases, the primary
mechanism dominates at low metallicities, while  the secondary one takes over when
the ISM becomes more metal--enriched. Nevertheless, despite that it seemed that the
old secondary--only GCR nucleosynthesis paradigm was unable to reproduce the Be and
B vs Fe slope, Fields \& Olive (1999) succeeded in doing so by using the recently
observed O/Fe relationship, with a single slope $ \simeq - $0.35 for [O/Fe] vs. [Fe/H]
at all metallicities (Israelian, Garc\'{\i}a L\'opez, \& Rebolo 1998; Boesgaard et
al. 1999a; Israelian et al. 2001). The situation about [O/Fe] evolution is, however,
far from being clear, as stated in Fulbright \& Kraft (1999). 

Most of the models for light element evolution published in recent years used
analytical approximations for the evolution of some of the important elements in the
LiBeB synthesis: carbon, nitrogen, oxygen, and iron (Fields \& Olive 1999; Fields et
al. 2000; Parizot \& Drury 2000; Ramaty et al. 2000). Some were closed
models (Vangioni-Flam et al 1998, Fields \& Olive 1999, Fields et al. 2000), which
are unable to solve the G--dwarf problem, as stated in Pagel (1997). 

Concerning lithium evolution, several works have recently been published,
involving different types of possible stellar sources (Romano et al. 1999), one of
which, massive AGB stars, seems to be recently ruled out as an important source
(Ventura, D'Antona, \& Mazzitelli 2000). Casuso \& Beckman (2000) have
proposed a model that only considers galactic cosmic ray nucleosynthesis, but with a
GCR rate proportional to the gas ejection rate from stars with masses lower than 3
M$_\sun$.

In this article we analyze the evolution of the light elements by means of a
detailed numerical model for the galactic chemical evolution (Alib\'es, Labay, \&
Canal 2001; hereafter ALC). Our model assumes that the Galaxy builds by infall of
extragalactic material, in an inside-out scenario for Galaxy formation,
and adequately  reproduces the main solar neighborhood observational constraints:
age--metallicity relation, G--dwarf metallicity distribution, current star formation
(SFR) and supernova rates, and the evolution of the main elements up to the iron
peak (see figures and tables in ALC).  In section \ref{model} the model is briefly
described and the results are presented in section \ref{results}. The conclusions
are given in section \ref{conclusions}.

\section{THE MODEL\label{model}}

We solve numerically the classical equations for the chemical evolution of the Galaxy, relaxing the instantaneous recycling approximation. The same ingredients as in ALC are used for our LiBeB evolution model. That includes the double infall of Chiappini, Matteucci, \& Gratton (1997) given by:
\begin{equation}
\label{infall}
\frac{d\sigma (t)}{dt}=A\ e^{-t/\tau _{\mathrm{T}}}+B\ e^{-(t-t_{\mathrm{max}})/\tau _{\mathrm{D}}}
\end{equation}
where $ \sigma (t) $ is the total surface mass density, $ \tau _{\mathrm{T}} $ and $ \tau_{\mathrm{D}} $ are the characteristic infall timescales for the halo--thick disk phase and for the thin disk phase, respectively, and $ t_{\mathrm{max}} $ is the time of maximum accretion into the thin disk. We take $ t_{\mathrm{max}}=\tau _{\mathrm{T}}  = 1$ Gyr and $ \tau _{\mathrm{D}} = 7$ Gyr. The coefficients $ A $ and $ B$ are fixed by imposing that they reproduce the current solar neighborhood values of the total and halo mass surface densities.

An enriched infall model is adopted, which accretes primordial matter during
the first Gyr, and then begins to incorporate slightly enriched material, with a
metallicity of Z = 0.1 Z$ _{\sun }$, as suggested by the recent observations by
Wakker et al. (1999) of a massive ($ \sim  $10$ ^{7} $ M$ _{\sun } $)
metal--enriched (0.09 Z$ _{\sun } $) cloud that is now falling into the disk. We have
also considered the more usual assumption of a  primordial composition for the whole
infall process. Although in the first case there is a small contribution to the
light element abundances from the infall, the final results for both types of
infall are almost indistinguishable, as it will be shown later.

The initial mass function (IMF), $\Phi$, used is that of Kroupa, Tout, \& Gilmore
(1993), which takes different slopes for M/M$ _{\sun }  \leq  0.5$, $0.5 \leq$
M/M$ _{\sun }  \leq  1.0$, and M/M$ _{\sun }  \geq  1.0$. The range of stellar
masses considered goes from 0.08 to 100 M$ _{\sun }$.

The star formation rate is from Dopita \& Ryder (1994) (eq. \ref{SFR}) with $ m =
5/3$, $ n = 1/3$ and $ \nu  = 1.2$:

\begin{equation} 
\label{SFR} \Psi (r,t)=\nu 
\frac{\sigma ^{n}(r,t)\sigma
_{\mathrm{g}}^{m}(r,t)}{\sigma ^{n+m-1}(r_{\sun },t)}\ \mathrm{M_{\sun } pc^{-2} Gyr^{-1}}
\end{equation}
applied to the solar galactocentric distance of 8.5 kpc.

\begin{figure}
\plotone{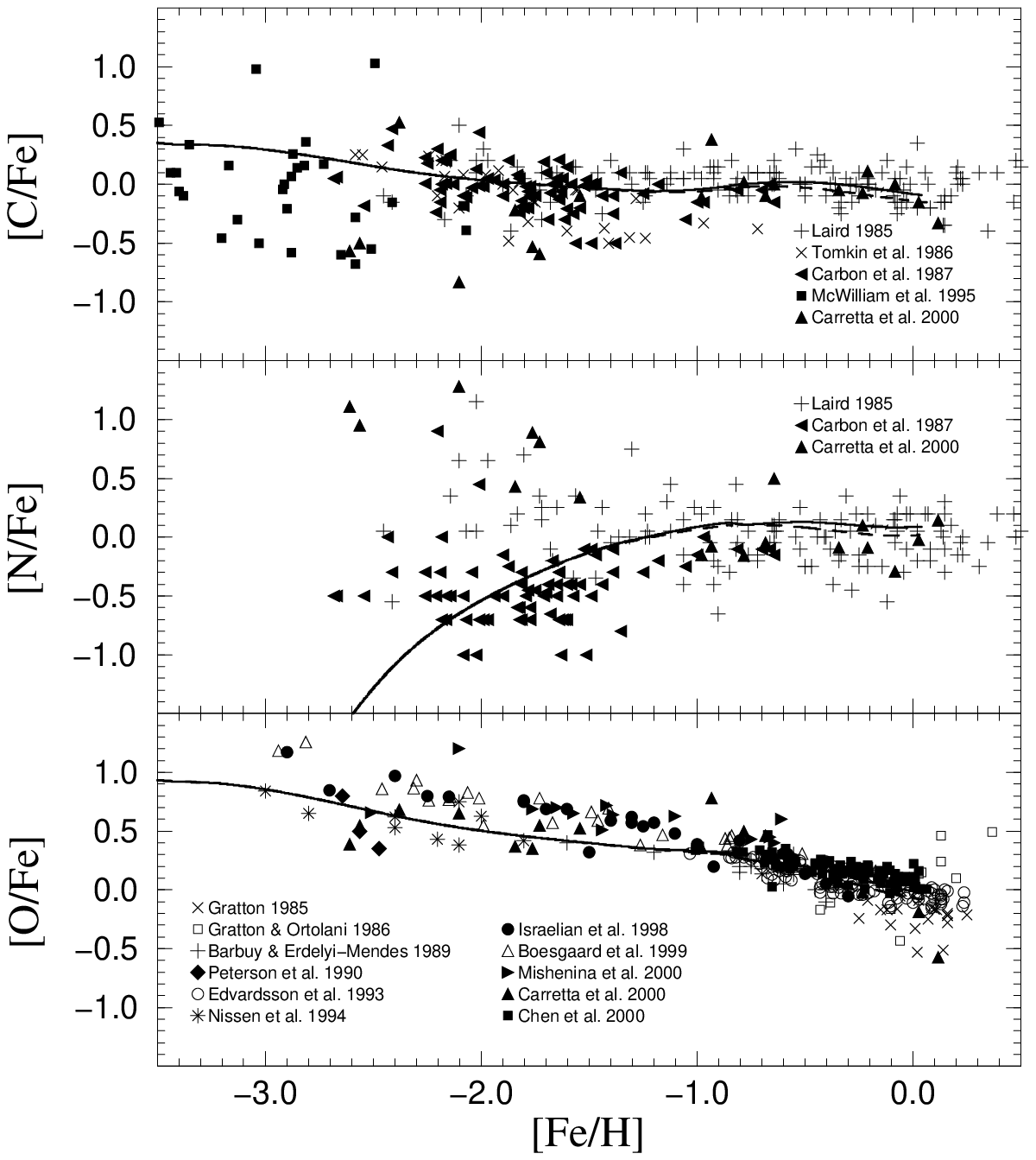}
\caption{Carbon, nitrogen, and oxygen evolution, from ALC. \emph{Solid line}:
enriched model; \emph{dashed line}: primordial model.}
\label{CNOgraf}
\end{figure}

To account for the CNO evolution some of the most recent and most extensive (both in
mass and metallicity) yield calculations are used: van den Hoek \& Groenewegen
(1997) for low-- and intermediate--mass stars, Woosley \& Weaver (1995)\footnote{It
is reminded here that, as in our previous solar neighborhood model (ALC), only half the iron yields calculated by these authors is adopted} for type II supernovae and
model W7 from Thielemann et al. (1996) for type Ia supernovae. Our results for the
CNO evolution are shown in Figure \ref{CNOgraf} for the two infall compositions.
Carbon and nitrogen follow the observational data, while [O/Fe] shows a gradual
increase towards low metallicity with a slope of $\sim -0.28$. This value is
slightly lower than the values obtained from observations of the \ion{O}{1} triplet
(Abia \& Rebolo 1989; Israelian et al. 1998, 2001; Boesgaard et al. 1999a) of
$\approx -0.35$.

\placefigure{CNOgraf}

\subsection{LiBeB Sources\label{sources}}

In Meneguzzi et al. (1971), only spallation reactions induced by GCR and Big Bang
nucleosynthesis of $ ^{7} $Li were considered as sources of LiBeB. Since then,
several stellar mechanisms have been proposed as sources of some of the light
elements: the $ \nu  $--process (possible solution to the solar system boron and
lithium isotopic ratios problem), nova outbursts (possible producers of $ ^{7} $Li)
and C--stars (identified producers of $ ^{7} $Li). All of them are included in our
evolution code. Massive AGB stars are not included as a source of $ ^{7} $Li,
since recent works (Ventura et al. 2000) insist on their small contribution to the
evolution of this isotope.

\subsubsection{Big Bang Nucleosynthesis}

$ ^{7} $Li is the only light element that is produced in a significant amount in 
standard big bang nucleosynthesis (BBN) models. In non--standard BBN, some amount of
other isotopes could also be produced, but the data available at low metallicities
do not indicate the presence of any kind of abundance plateau for them. A BBN
contribution of [$ ^{7} $Li] = 2.2\footnote{[X]=12$ +\log \left(
\frac{X}{H}\right)$} has been included according to Molaro (1999), which ensures the
existence of the lithium or Spite plateau (Spite \& Spite 1982) up to [Fe/H]$
\approx - $1.5. {Due to this BBN production of $ ^{7} $Li, open and closed box
models of Li evolution do not give the same results, since in closed box models all the
primordial lithium is already in place when the evolution begins. Therefore, closed
box models with the same nucleosynthetic prescriptions than open ones do produce an
overabundance of Li at [Fe/H] = 0 by a factor $\sim 1.6$. This is not the case for
Be and B, for which open and closed box models give similar results.}

\subsubsection{Galactic Cosmic Ray Nucleosynthesis}

As in Parizot \& Drury (2000), we adopt a composition of the Galactic Cosmic Rays ($
X_{i,\mathrm{GCR}} $) corresponding to their being accelerated inside a superbubble,
where newly synthesized material ejected by a supernova ($ X_{i,\mathrm{ej}} $) is
accelerated by the shock waves generated by other SN and mixed with the ISM of that epoch ($
X_{i,\mathrm{ISM}} $), the evolution of the latter being that shown in Figure
\ref{CNOgraf}. 

\begin{equation} 
\label{CR-comp} 
X_{i,\mathrm{GCR}}(t)=\alpha _{\mathrm{ej}}X_{i,\mathrm{ej}}(t)+(1-\alpha
_{\mathrm{ej}})X_{i,\mathrm{ISM}}(t) 
\end{equation} 

Different compositions of the GCR are used by changing the free parameter $
\alpha _{\mathrm{ej}} $ in eq. \ref{CR-comp}. For $ \alpha _{\mathrm{ej}}  = 0$ we
have pure ISM CR composition and for $ \alpha _{\mathrm{ej}}  =$ 1, pure SNII ejecta
CR composition.

The CR source spectrum considered is the so--called shock acceleration spectrum:
\begin{equation}
\label{CRspec}
q(E)\propto \frac{p^{-s}}{\beta }e^{-E/E_{0}}
\end{equation}
where $ p $ and $ E $ are the particle momentum and kinetic energy, both
per nucleon, $ \beta = v/c $ and, for the two parameters, we have
chosen $ s  = 2.2$ and $ E_{0}  = 10$ GeV/n. $ E_{0} $ takes into account
the possibility that the spectrum is cut off at high energies. Since a leaky box model for GCR propagation is adopted, a escape path of CR from the Galaxy of
$ X_{esc}  =$ 10 g cm$ ^{-2} $ has been included in the propagation calculations. 

For the calculation of the  production rate by spallation reactions we have used a
revised version of the \texttt{LiBeB} code from Ramaty et al. (1997), which includes
the cross sections of Read \& Viola (1984) and takes into account all the possible
reactions leading to any LiBeB isotope, both direct and inverse reactions (including
two--step processes like $ ^{16} $O($ p $,$ x $)$ ^{12} $C($ p $,$ x' $)$ ^{7} $Li).
Since  GCR particles are supposed to come from the material inside  a SB, the GCR
flux, and therefore the spallation reaction production rate, is made proportional
to  the Type II supernova rate.

We do not consider any low--energy cosmic rays (LECR), since the only observational
evidence of this type of CR, spotted in the Orion star formation region, has been
withdrawn (Bloemen et al. 1999). Future X--ray and gamma--ray line observations
should clarify whether LECR actually exist.

\subsubsection{The $ \nu  $--process}

Woosley \& Weaver (1995), in their calculations of the yields from type II
supernovae, took into account the contribution from the neutrino--induced
nucleosynthesis, i.e. the spallation reactions between the huge flux of neutrinos
ejected by the explosion and the material, newly synthesized, in the intermediate
layers of the star. This primary mechanism, which has not been observationally
corroborated, produces mainly $ ^{7} $Li and $ ^{11} $B, so it could be quite
important because GCR nucleosynthesis alone can not reproduce the Solar System
lithium and boron isotopic ratios.  Since these yields are quite uncertain because
the flux intensity and energy spectrum of the neutrinos are not well known, a
correction factor, $ f_{\nu } $, is introduced, its value being obtained by  fitting
the meteoritic boron isotopic ratio.

\subsubsection{Novae}

Nova outbursts could be important nucleosynthetic sites for those elements which
have overproduction factors there, relative to solar, above 1000. According to the
recent yields for CO and ONe nova outbursts from Jos\'e \& Hernanz (1998), that is
the case for $ ^{7} $Li (nova outbursts  produce, during the thermonuclear runaway,
$ ^{7} $Be, which decays into $ ^{7} $Li). We have, thus, adopted a $ ^{7} $Li
average yield per nova outburst of 1.03$ \times  $10$ ^{-10} $ M$ _{\sun } $, taking
into account that 30\% of the nova outbursts come from ONe white dwarfs, and an
outburst rate given by the following equation:

\begin{eqnarray}  
\frac{dR_{outburts}}{dt} & = & D\int_{M(t)+0.5}^{9.5}\frac{\Phi
(M_{B})}{M_{B}}dM_{B} \nonumber \\ & & \int _{\mu _{m}}^{\mu _{M}}f(\mu )\Psi
(t-\tau _{M_{B}(1-\mu )}-t_{\mathrm{cool}})d\mu    
\end{eqnarray}  
where $ D $
ensures that the current nova outburst rate is $ \sim  $40 yr$ ^{-1} $ (Hatano et
al. 1997), $ M_{B} $ is the binary mass, $ \mu  $ is the ratio of the mass of the
secondary star to that of the whole binary system and finally $ f(\mu ) $ is the
binary mass--ratio distribution function of Greggio \& Renzini (1993). The time for the WD to cool enough to be able to start producing outbursts, $t_{\mathrm{cool}}$, is set to 1 Gyr. $^{7} $Li production by nova outbursts could be confirmed by the future INTEGRAL
mission.

\subsubsection{C--stars}

Even if supernovae ($ \nu  $--process) and novae are theoretically possible sources
of $ ^{7} $Li, carbon--rich stars (C--stars) with C/O $> 1$ give the only
undeniable evidence for an stellar origin of this isotope, since they are
seen to be Li--rich. The majority of them show low carbon isotopic ratios 
($^{12} $C/$ ^{13} $C$ < 15$) and are of J type (Abia \& Isern 1997), with masses
lower that 2--3 M$_{\sun}$.  The mechanism by which $ ^{7} $Li is produced and
ejected to the ISM is not well established, so the only way to estimate the yield is
empirical. Abia, Isern, \& Canal (1993) statistically analyzed the contribution of
C--stars to the $ ^{7} $Li nucleosynthesis and suggested a time--dependent
production rate:

\begin{equation}
\label{rCstars}
S_{7}(t)=S_{7}^{\sun }\frac{\int _{M_{\mathrm{l}}}^{M_{\mathrm{u}}}\Phi (M)\Psi (t-\tau _{M})dM}{\int _{M_{\mathrm{l}}}^{M_{\mathrm{u}}}\Phi (M)\Psi (t_{G}-\tau _{M})dM}
\end{equation}
where $ M_{\mathrm{l}}  = 1.2$ M$ _{\sun } $ and $ M_{\mathrm{u}}  = 3$ M$ _{\sun } $. They
estimated empirically a current production rate, from a sample of galactic C--stars,
of $ S_{7}^{\sun }  = 2 \times  10 ^{-9} $ M$ _{\sun } $pc$ ^{-2} $Gyr$ ^{-1} $.
The authors point out that this  $^{7} $Li production rate can only be a lower limit and they needed to increase it to $6 \times  10^{-8} $ M$ _{\sun } $pc$ ^{-2} $Gyr$ ^{-1} $ (Abia, Isern \& Canal 1995) in order to obtain a good fit to the solar system data. In the present model, this rate has only to be increased up to $1.5 \times  10^{-8} $ M$_{\sun } $pc$ ^{-2} $Gyr$ ^{-1} $.

\subsection{Destruction of LiBeB}

\begin{figure}
\plotone{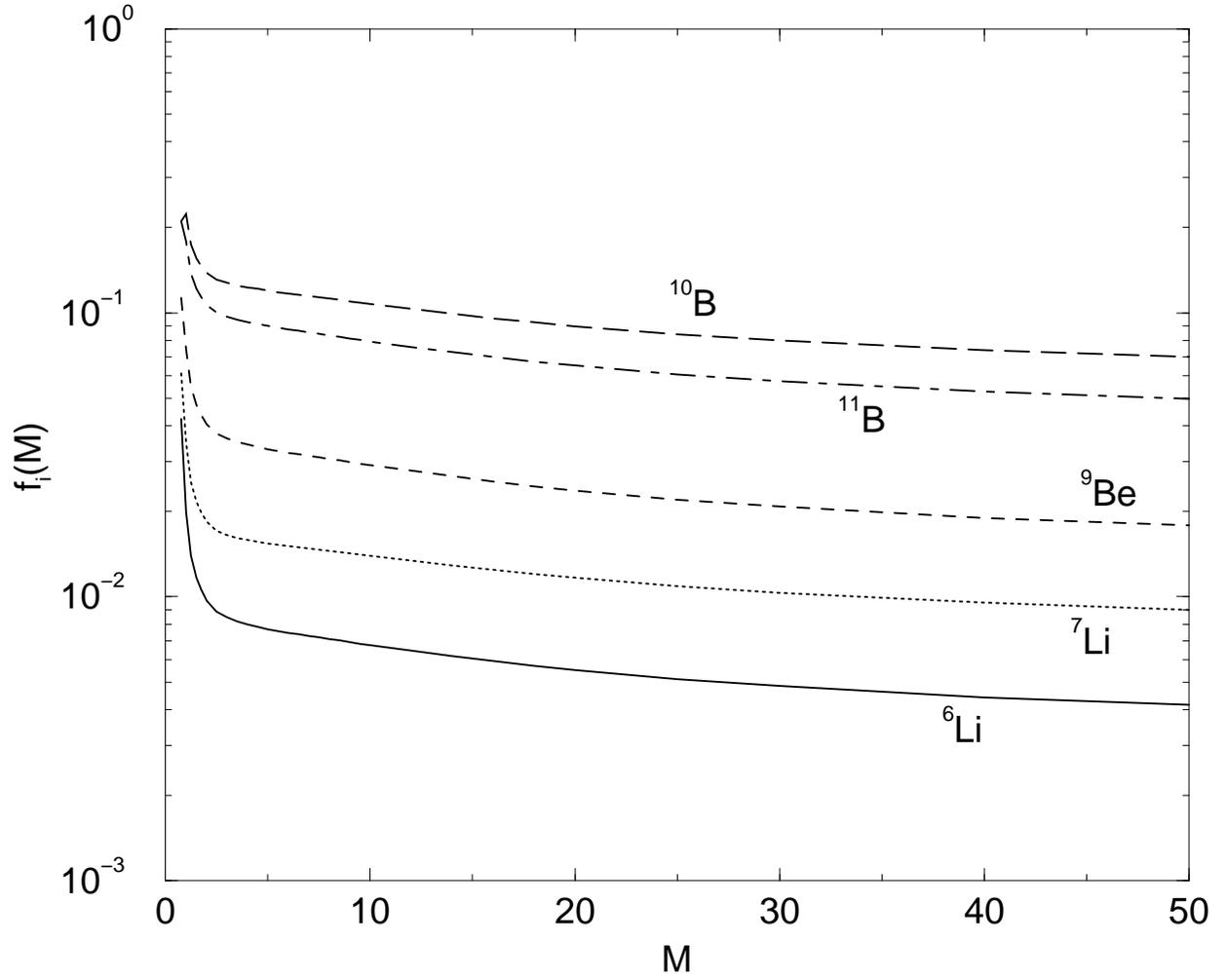}
\caption{Astration model results. Mass fraction of a LiBeB isotope
that has survived and has been ejected from each star ($ f_{i}(M) $).}
\label{astration}
\end{figure}

Traditionally, LiBeB isotopes are supposed to be completely destroyed in stars, so
when they die they do not contribute to the ISM abundance. This assumption has been
checked by means of main--sequence stellar models (Hansen \& Kawaler 1994) that
include all the nuclear reactions involved in LiBeB destruction, whose rates are
taken from Caughlan \& Fowler (1988). We have evaluated, for the whole range of
stellar masses, the fraction of each LiBeB isotope that survives and is ejected from
the star. Our results are displayed in Figure \ref{astration}. As expected, the two
lithium isotopes are the most fragile ones within the LiBeB group, and they are
almost completely destroyed in stars over a wide range of stellar masses.  Beryllium
and both boron isotopes are more resistant, since they are destroyed at higher
temperatures, but nevertheless the fraction of the original BeB that survives is, as
in the case of the lithium isotopes, very small.

These destruction terms are included in our calculations, and there are only minute
differences in the final results as compared with those obtained assuming complete
destruction.

\placefigure{astration}

\section{\label{results}RESULTS}

We have analyzed the evolution of the light elements, taking into account all the
sources described in section \ref{sources} and the restrictions that energetics
puts  on spectrum and composition of GCR. {In Table \ref{tablesources} a summary of all the different sources considered is presented. It is important to notice that no other combination of the contribution from the different sources could reproduce all the constraints considered (Li plateau, beryllium evolution versus iron, and solar system boron isotopic ratio and lithium abundance).}

\begin{table}
\begin{center}

\caption{SUMMARY OF THE DIFFERENT SOURCES CONSIDERED}
\label{tablesources}
\begin{tabular}{lcll}
\hline
\hline
source&isotopes produced&contribution&fixed by\\
\hline 
BBN&$^7$Li&[$^7$Li]=2.2&Li plateau\\
GCRN&LiBeB&$ \alpha _{\mathrm{ej}}=0.25 $&Be vs Fe\\
$\nu$--process&$^7$Li and $^{11}$B&$f_{\nu}=0.29$&($^{11}$B/$^{10}$B)$_{\sun}$\\
novae&$^7$Li&averaged yield&--\\
C--stars&$^7$Li&$S^{\sun}_7=1.5 \times  10^{-8} $ M$
_{\sun } $pc$ ^{-2} $Gyr$ ^{-1} $&Li$_{\sun}$\\
\hline
\end{tabular}

\end{center}
\end{table}

\subsection{Abundances Evolution}

\begin{figure}
\plottwo{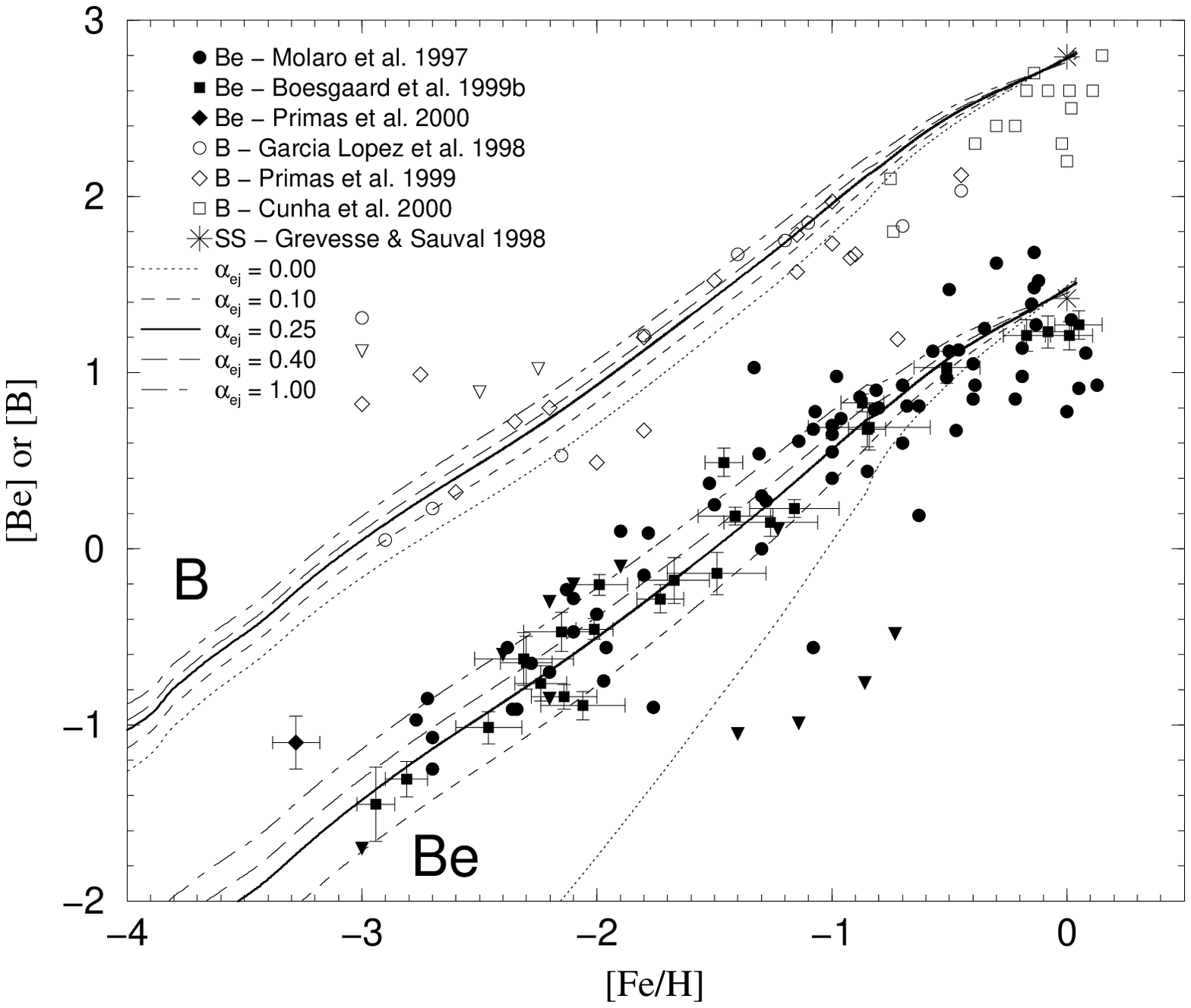}{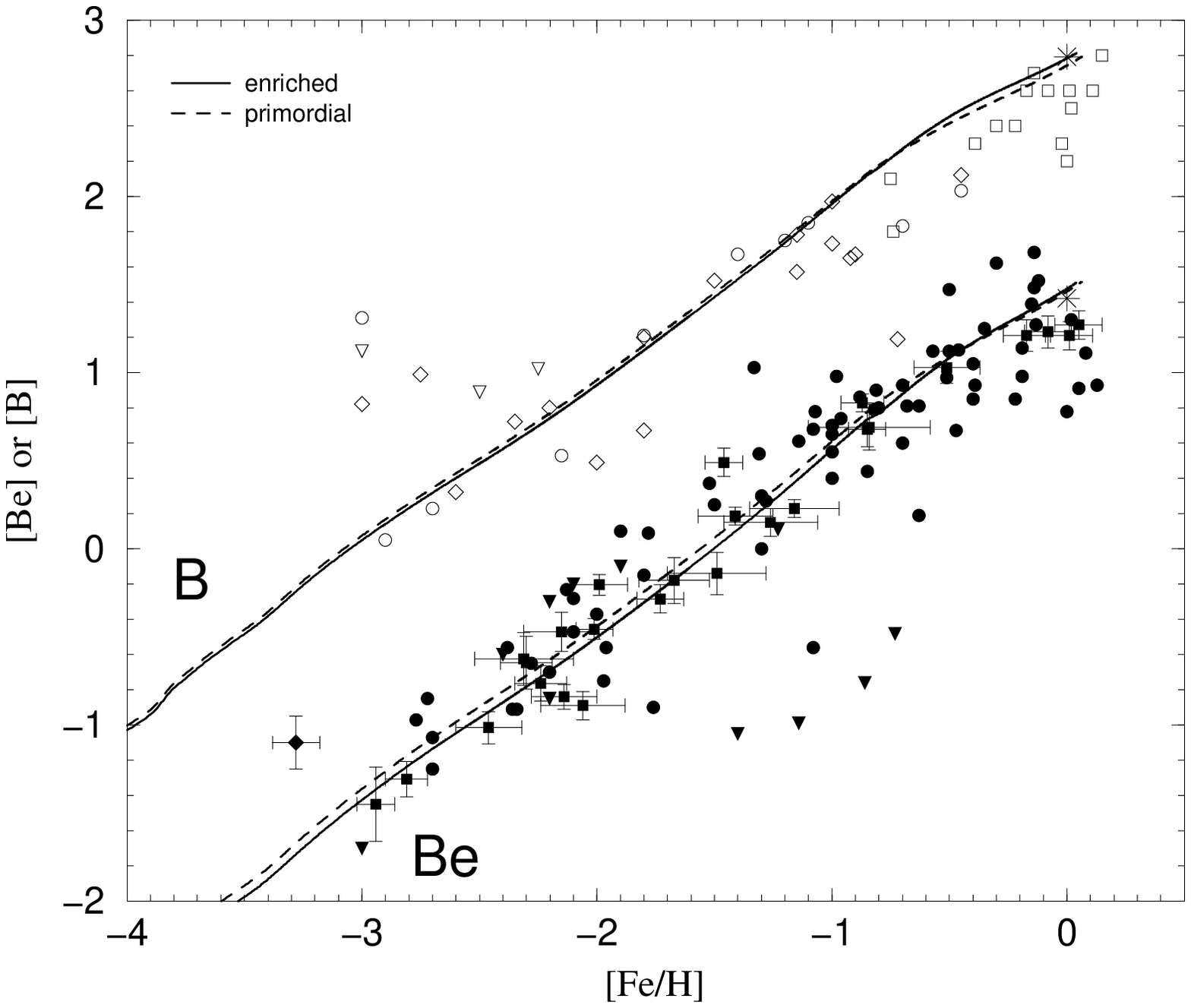}
\caption{Beryllium and boron evolution. \emph{Left panel}: evolution for the
enriched model and different values of $\alpha_{\mathrm{ej}}$. \emph{Right panel}: evolution
for the $\alpha_{\mathrm{ej}}=0.25$ model and the two infall compositions.}
\label{BeB.mix}
\end{figure}

Concerning the GCR composition, we show in the left panel of Figure \ref{BeB.mix}
that for values of $ \alpha _{\mathrm{ej}} $ (see eq. \ref{CR-comp}) near 0.25
(solid line) good fits to the observed Be and B evolution as a function of [Fe/H]
are obtained. {The $\chi^{2}$ test gives an interval for the possible $ \alpha
_{\mathrm{ej}} $ values of (0.12,0.41) for a level of confidence of 90\%.} For
values of $ \alpha _{\mathrm{ej}} $ near 0 (GCR made only from accelerated ISM), a
slope of 2 results, while, if taking a value of $ \alpha _{\mathrm{ej}} $ close to 1
(GCR made of SNe ejecta) the contribution from GCR nucleosynthesis to LiBeB
evolution at low metallicities is too large. The preceding means that a double origin for the composition
of GCR must be considered, since  both a pure SNII composition and a
pure ISM composition can be rejected. Such a  mixture is not well explained in non--magnetic SB models
(which favour $ \alpha _{\mathrm{ej}}\approx 0.02 $), unless we assume that the bulk
of the GCR comes from the central part of the SB, where the SN ejecta material would
be more dominant (Higdon et al. 1998). However, magnetic SB models (Tomisaka 1992)
give higher values for $ \alpha _{\mathrm{ej}} $, because magnetic fields reduce
heat conduction, and thus the evaporation of the ISM shell surrounding the SB. The
value of $ \alpha _{\mathrm{ej}} $ for these latter models is close to our best fit. In the
right panel of Figure \ref{BeB.mix}, we compare the evolution, in the $ \alpha
_{\mathrm{ej}}=0.25$ case, for the two different infall compositions. One sees that
the evolution is practically unaffected by the chemical composition of the accreted
matter, as long as its enrichment remains moderate. 

The very recent beryllium abundance determination in the very metal--poor halo star
G 64--12 ([Fe/H]= $-$3.30 and [Be]= $-$1.1) by Primas et al. (2000) is not well
fitted. As stated by the authors, this value would either suggest a flattening of
the Be vs Fe relationship at low metallicities or dispersion of values in the early
Galaxy. More low--metallicity data are needed to confirm either of these suggestions.

\placefigure{BeB.mix}

{Parizot (2000), also within the framework of a SB model for the acceleration of
GCR, analyzed the proportion of pure ejecta of supernovae accelerated in
superbubbles (our parameter $ \alpha _{\mathrm{ej}} $). He found a best value of $
0.03 $ when adopting a GCR energy spectrum $ q(E)\propto E^{-1}\exp(-E/E_{0}) $ with
a rather low cut-off energy of $ E_{0}=500 $ MeV/n, that is supposed to mimic the
results of particle acceleration in SB by a collection of weak shocks. Thus, our
best value of $ \alpha _{\mathrm{ej}}=0.25 $ is a factor of $ \sim 8 $ larger than
the result obtained by Parizot (2000). Such discrepancy comes in part from the
higher efficiency of his weak SB spectrum. Had  this spectrum been used instead of
that in eq. (4), a larger Be production per erg would have been obtained, by a
factor of $ \sim 2.2 $,  thus reducing by the same factor our mixing coefficient $
\alpha _{\mathrm{ej}} $. A further factor of $ \sim 3 $ is due to the fact that
Parizot (2000) did not calculate the evolution of the actual Be/O abundance ratio,
but only that of the Be/O production ratio by dividing the yields of Be and O per
supernova. In that way, stellar astration, which is important for late times and
high metallicities, was not taken into account. Astration severely reduces the
abundance of Be, but not that of O, thus lowering the Be/O abundance ratio at late
times (Fields et al. 2001). Our model, neglecting stellar astration, produces a Be
abundance at {[}Fe/H{]}$ =0 $ which is $ \sim 3 $ times larger than the value
measured in the Solar System. Finally, the study by Parizot (2000) does not follow
the detailed evolution of the composition of the ISM by means of a chemical
evolution model. Instead, he assumes solar composition scaled with metallicity,
i.e. O/H, which translates into overestimated C and N abundances, both in the ISM and
in the $ 1-\alpha_{\mathrm{ej}} $ fraction of GCR accelerated from the ISM.}

\begin{figure}
\plotone{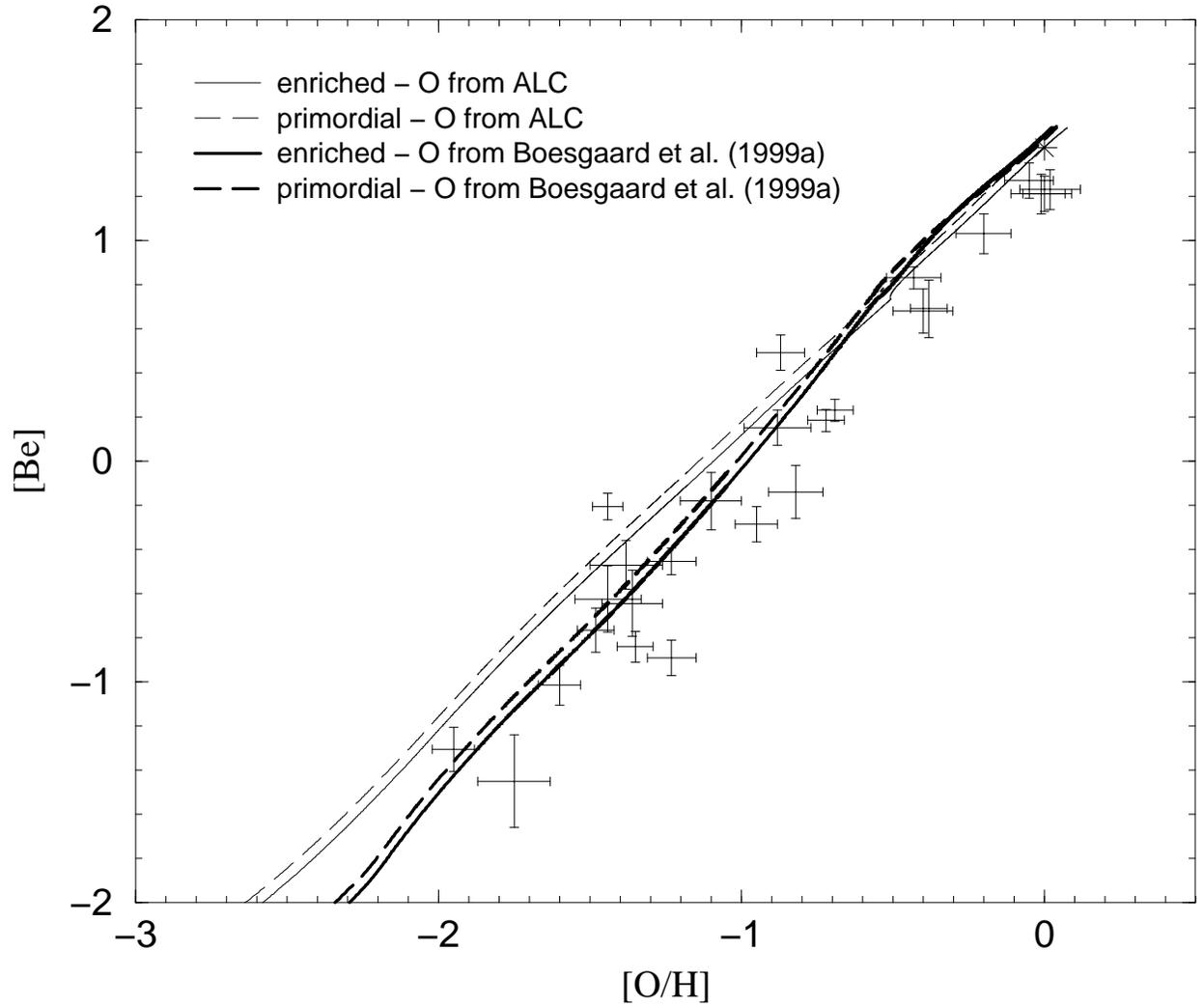}
\caption{Beryllium vs oxygen evolution for the $\alpha_{\mathrm{ej}}=0.25$ case. Oxygen resulting from our model is used in
\emph{thin line} plots, while oxygen from Boesgaard et al. (1999a) is used in the \emph{thick line} ones.
Data: Boesgaard et al. (1999b).}
\label{BevsO}
\end{figure}

Since beryllium is supposed to be produced only by GCR, and therefore its abundance 
is expected to be directly dependent on the oxygen abundance in the ISM, it is
interesting to compare our model results with the observational data in a [Be] vs
[O/H] plot (Fig. \ref{BevsO}). We notice that if we use the oxygen calculated in
ALC, the best value of the parameter $ \alpha _{\mathrm{ej}} $ does not seem to be
the same as that obtained from the [Be] vs [Fe/H] plot. This flaw was expected since
our calculated oxygen evolution does not perfectly fit the Israelian et al. (1998,
2001) and the Boesgaard et al. (1999a) data, although our results are closer to
them than those obtained in other recently published galactic chemical evolution
models. However, when  their relationship for [O/Fe] vs [Fe/H] is used to transform
the Be-Fe plot to a Be-O one, a reasonable fit is recovered for $ \alpha
_{\mathrm{ej}}  = 0.25$, at least for the low metallicity region where our
results closely follow the mean trend of the data. At high metallicity we slightly
overpredict the Boesgaard et al. (1999b) data, but the Solar System value is well
reproduced still.  

\placefigure{BevsO}

\begin{figure}
\plottwo{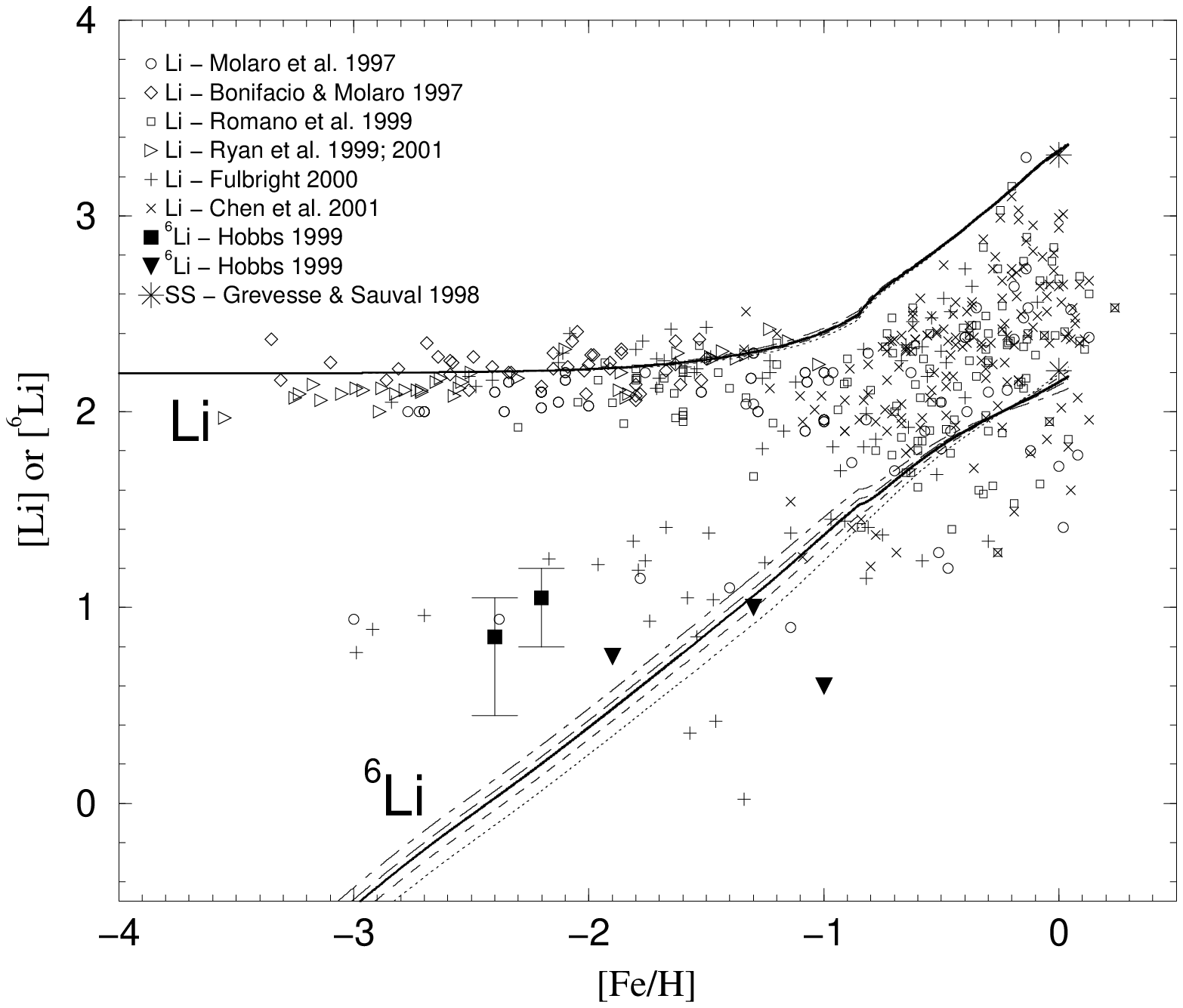}{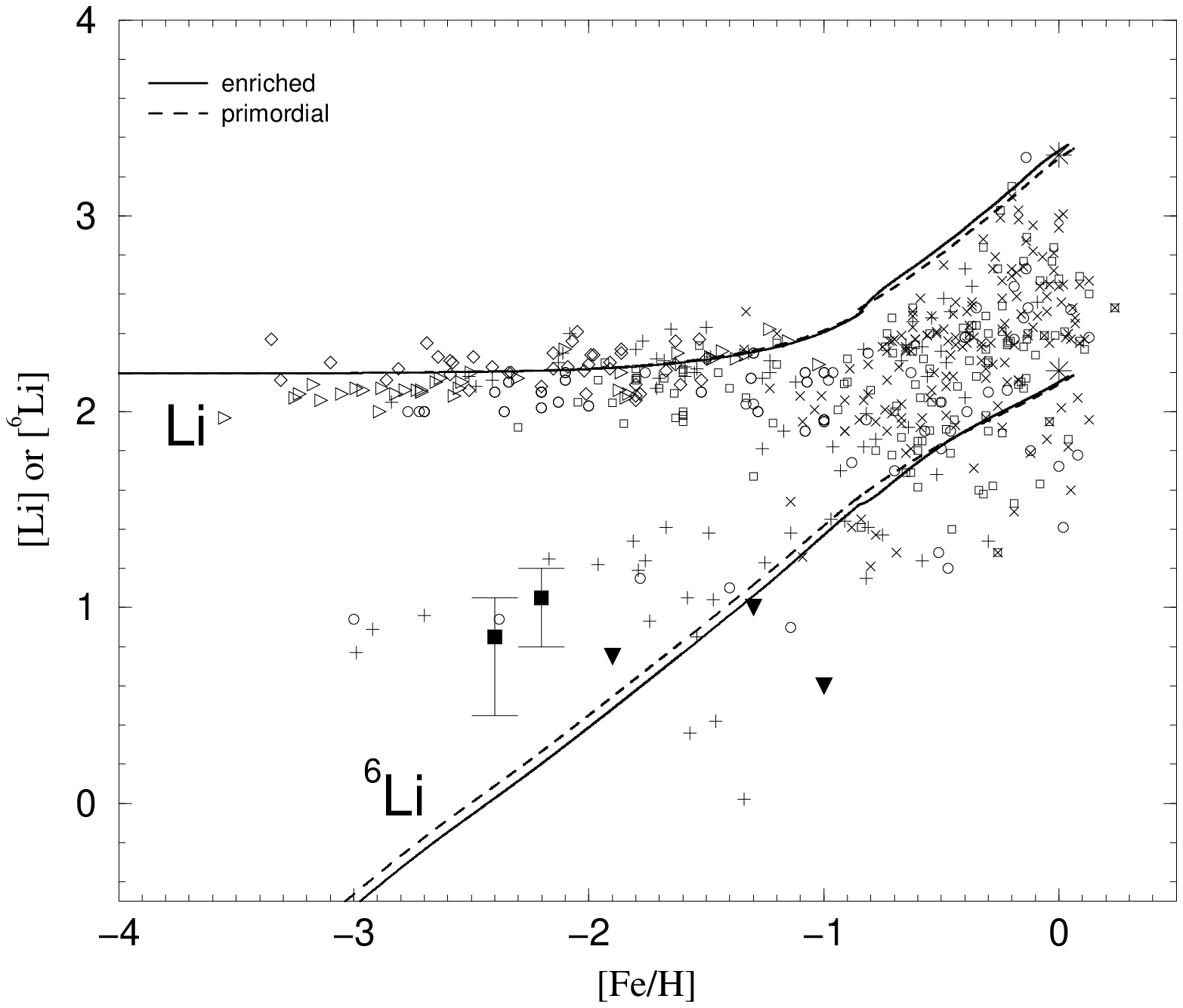}
\caption{Lithium and $ ^{6} $Li evolution. \emph{Left panel}: enriched model and same values of $ \alpha _{\mathrm{ej}} $ as in figure \ref{BeB.mix}. \emph{Right panel}: both infall models for $\alpha_{\mathrm{ej}}=0.25$. }
\label{li6li.mix}
\end{figure}

\begin{figure}
\plotone{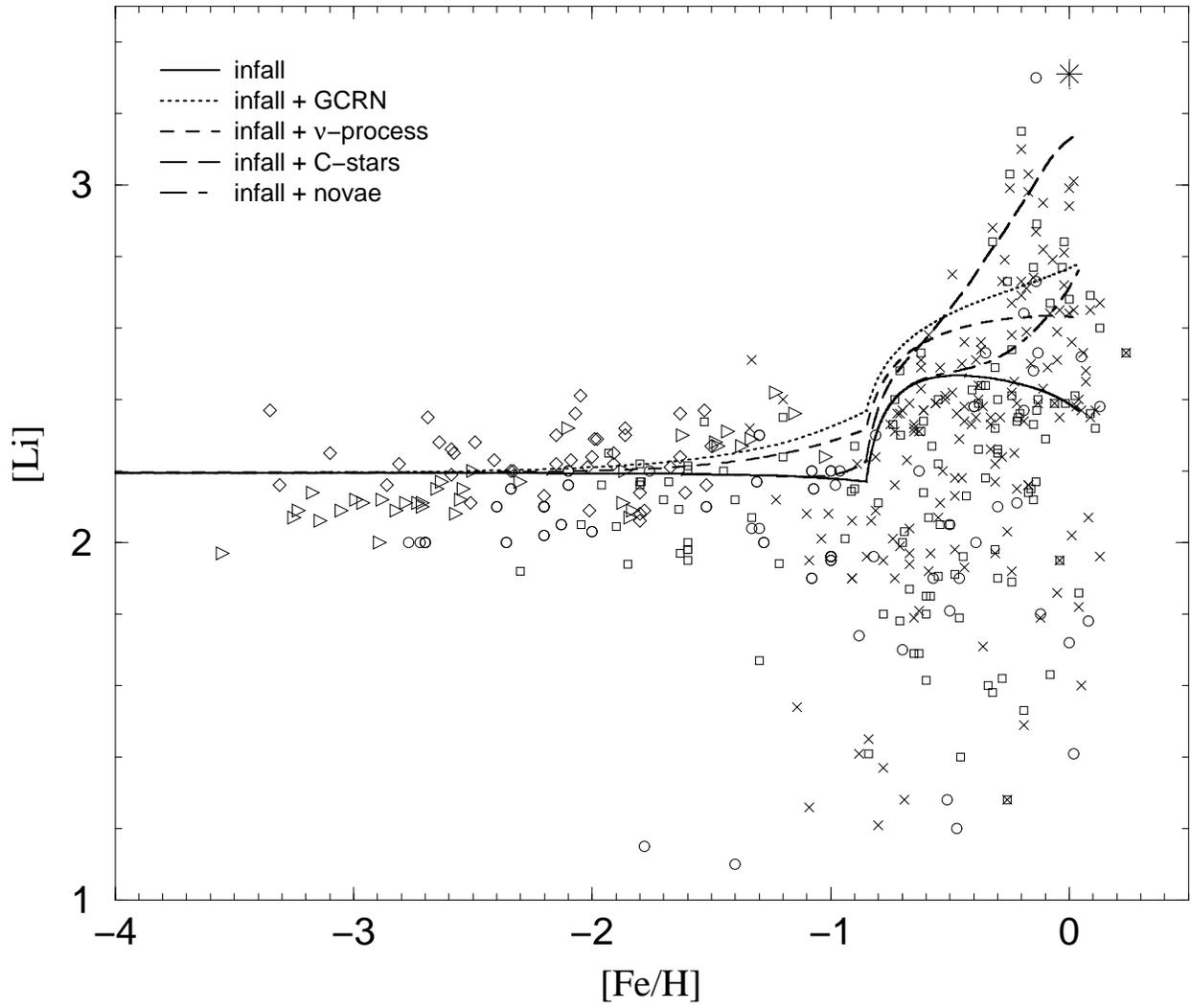}
\caption{Different contributions for the lithium evolution, for the enriched infall composition and $ \alpha _{\mathrm{ej}} = 0.25$.}
\label{licontrib}
\end{figure}

In Figure \ref{li6li.mix},  the evolution of total lithium and $ ^{6} $Li are shown:
in the left panel for several values of the parameter $ \alpha _{\mathrm{ej}}$ and
in the right panel for the two infall compositions and fixed $ \alpha
_{\mathrm{ej}}$. The lithium plateau is well reproduced, as well as the upper
envelope of the population I stars and the Solar System value. {As shown in
Figure \ref{licontrib},} the increase of the lithium abundance for [Fe/H] $> - $1.5
is due to the contribution from the neutrino--induced nucleosynthesis, and continues
by the additional Li from C--stars and  nova outbursts, whose contribution is
delayed to intermediate and high metallicities due to the long lifetimes of the
low--mass progenitors of those objects. {C--stars are the main contributors at
high metallicities, as it is clearly seen in the Figure}. The parameter $ \alpha
_{\mathrm{ej}} $ has little effect on the lithium evolution, because at low
metallicities, where changes of this parameter make the GCR nucleosynthesis
production rate change, the Li from the Big Bang and from the $\alpha + \alpha$ reactions
dominate. In the case of $ ^{6} $Li, the less abundant lithium isotope and still not
sufficiently measured at low metallicities, our model  reproduces the Solar System
value, but not the low--metallicity observations. As indicated by Ramaty et al.
(2000), this may suggest the need for an extra source of $ ^{6} $Li in the Galaxy. A
possible solution to this defect of $ ^{6} $Li production would be a higher
primordial value (some non--standard BBN models produce large $ ^{6} $Li abundances:
Jedamzik \& Rehm 2001), but that would mean the existence of   plateaus at low
metallicities also for Be and B (or a $ ^{7} $Li primordial value not compatible
with the Spite plateau), which have not been observed up to now.

As it happens for Be and B, the Li evolution is almost independent of the assumed composition of the infalling matter.

\placefigure{li6li.mix}

\placefigure{licontrib}

\subsection{Ratios Evolution}

Isotopic and elemental ratios are  crucial to analyze the importance of the
different contributions to LiBeB abundances. The boron isotopic ratio shows, for a
given value of $ \alpha _{\mathrm{ej}} $, the importance of the neutrino induced
nucleosynthesis. This source is necessary to reach the solar system value. On the
other hand, the boron over beryllium ratio data are quite constant at all
metallicities, implying that the $ \nu  $--process can not be dominant. The
data on the next element ratio considered, Li/B, show a steep descent from the large
value at low metallicities, due to the Li plateau, down to the solar value. Finally,
there are quite few available data for $ ^{6} $Li, so the lithium isotopic data are
not a strong constraint, the solar system data excepted.

\begin{figure}
\plotone{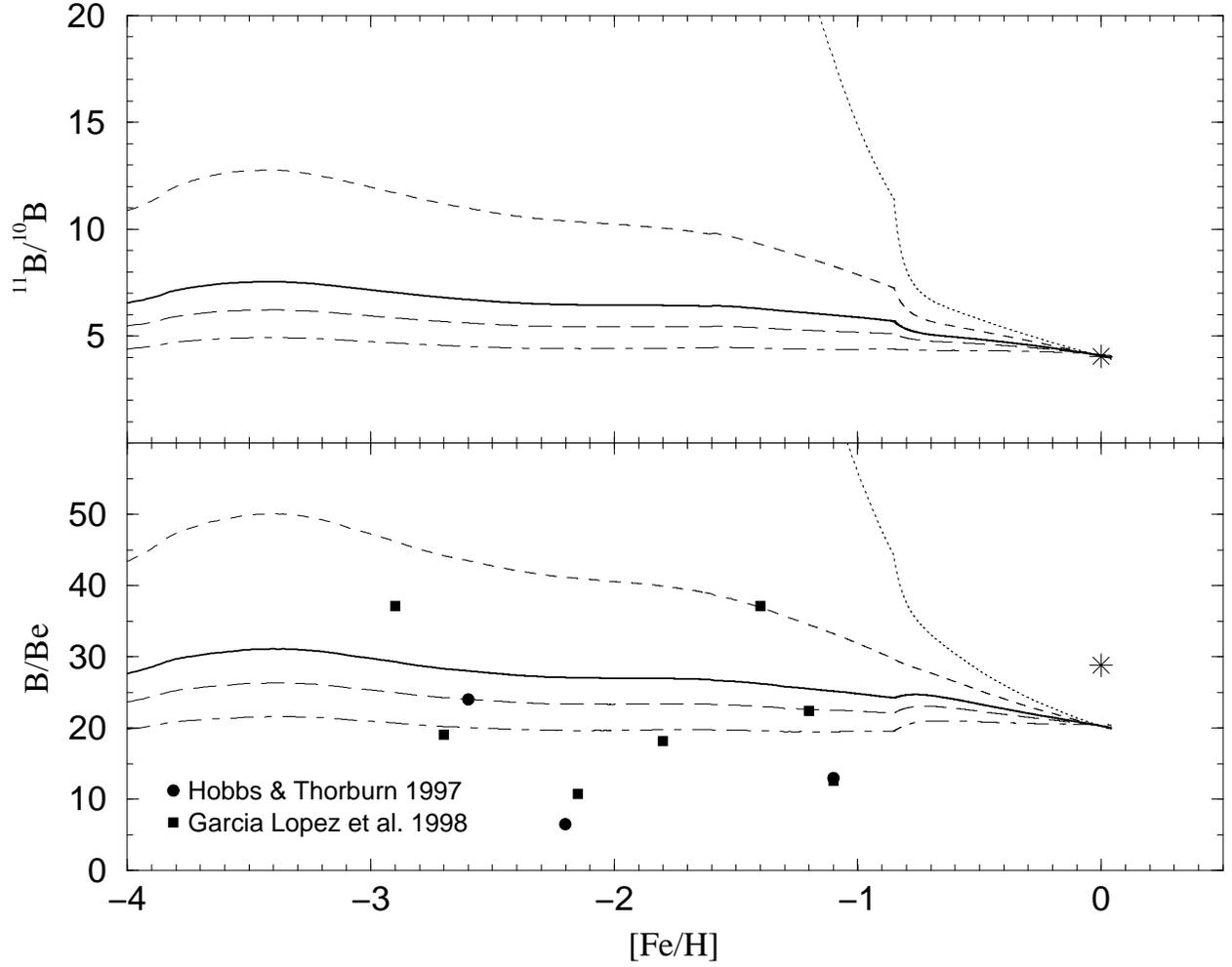}
\caption{$^{11}$B/$^{10}$B and B/Be ratios evolution. Same values of $ \alpha _{\mathrm{ej}} $ as in figure \ref{BeB.mix}, for the enriched infall composition. }
\label{1110-bbe}
\end{figure}

In all cases we have used a value of the parameter that restricts the contribution
of the neutrino induced nucleosynthesis of $ f_{\nu} = 0.29$ (similar to that found
by Ramaty et al. 2000), which allows to reproduce the Solar System boron isotopic
ratio, as displayed in the upper panel of Figure \ref{1110-bbe}.  Higher values of $
f_{\nu } $ would increase the ratio $ ^{11} $B/$ ^{10} $B up to values not allowed
by the Solar System data. We see that the ratio is quite constant during the halo
phase (a higher ratio for lower values of $ \alpha _{\mathrm{ej}} $) due to the
contribution of the $ \nu  $--process and the primary character of GCR
nucleosynthesis in those early epochs, and it decreases towards the Solar System
value when the secondary behaviour of the GCR becomes important enough. The lack of
data at low metallicity hinders a more firm conclusion, but the descending trend
after [Fe/H] = 0 is compatible with the current ISM ratio (3.4$ \pm  $0.7, Lambert
et al. 1998).

A similar behavior is shown by the B/Be ratio (lower panel of Fig. \ref{1110-bbe}).
The constant ratio is higher for lower values of $ \alpha _{\mathrm{ej}} $, and the standard GCR nucleosynthesis also dominates at high metallicity.

\placefigure{1110-bbe}

\begin{figure}
\plotone{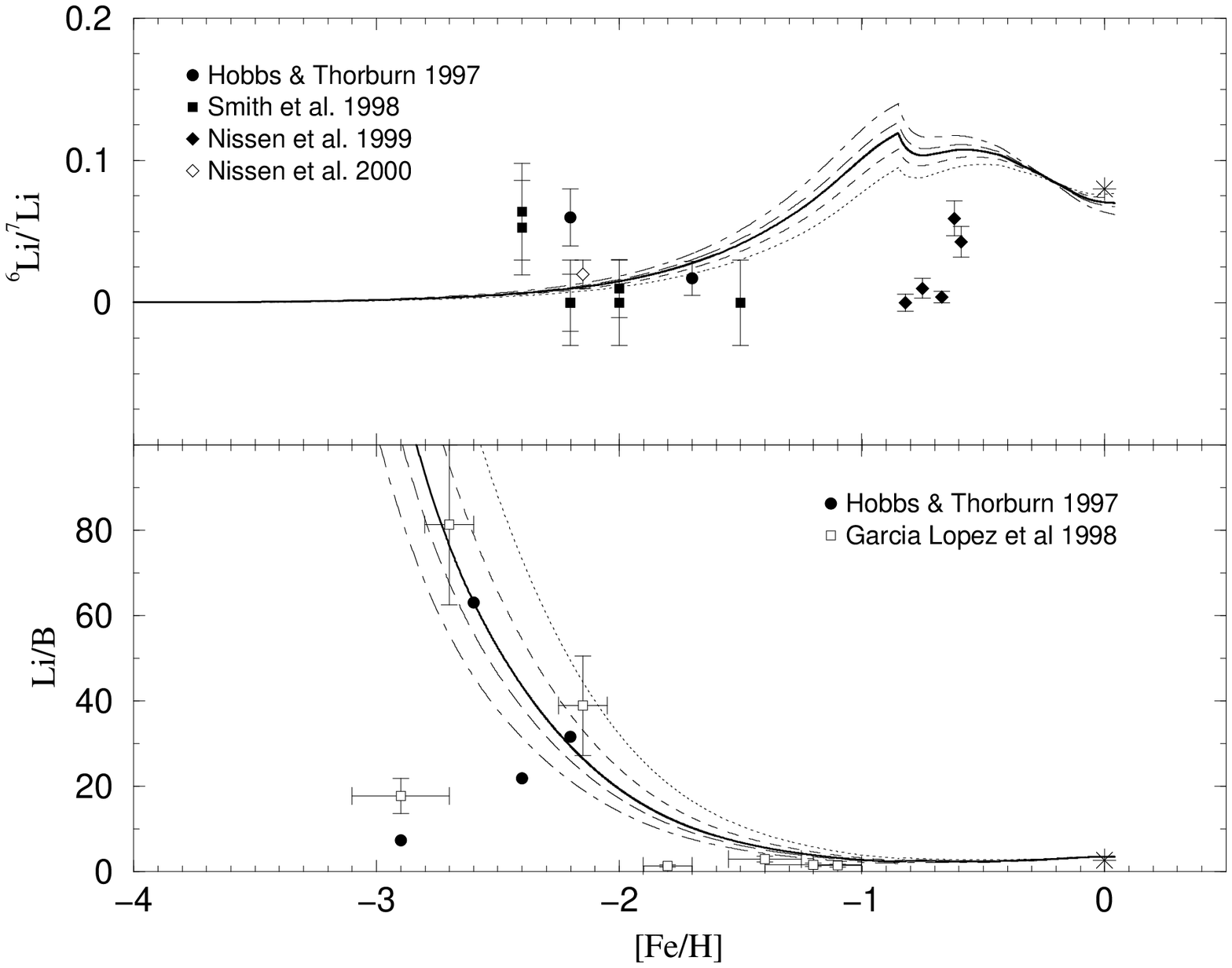}
\caption{$^{6}$Li/$^{7}$Li and Li/B ratio evolution. Same values of $ \alpha _{\mathrm{ej}} $
as in figure \ref{BeB.mix}, for the enriched infall composition. }
\label{67-lib.mix}
\end{figure}

Finally, the two other ratios considered include lithium (Figure \ref{67-lib.mix},
with $ ^{6} $Li/$ ^{7} $Li in the upper panel and Li/B in the lower one). As we have
already pointed out, our model can not account for  the $ ^{6} $Li at low
metallicities, so the lithium isotopic ratio obtained {is much lower than  the
nonzero} $^{6} $Li/$ ^{7}$Li  data {around [Fe/H] $=-2.5$}. {The values at
intermediate metallicities are not reproduced, since the stars where they have been
measured may have suffered some amount of lithium destruction (more important in the
case of $ ^{6} $Li than for $ ^{7} $Li).} For the Li/B ratio this lack of $ ^{6} $Li
is not important and our models fit well the available data.

\placefigure{67-lib.mix}

\subsection{Energetics}

Since supernovae, either individually or in SB, are supposed to be the site of GCR
acceleration, the total energy of a SN explosion must be larger than the energy of
the GCR particles produced by it. The mean value of the Be/Fe ratio in low--metallicity stars
measured by Molaro et al. (1997) and Boesgaard et al. (1999b) is about $1.37 \times 
10^{-6} $ and the average yield of $ ^{56}$Fe from SN, according to ALC prescriptions,
is $Q_{\mathrm{SN}} (\mathrm{Fe}) =  0.0515$ M$ _{\sun }\equiv 1.1\times 10^{54} $
atoms. Then, as shown by Ramaty et al. (1997), one can evaluate the number of atoms
of Be produced by each SN event:

\begin{equation}
Q_{\mathrm{SN}}(\mathrm{Be})=\frac{\mathrm{Be}}{\mathrm{Fe}}Q_{\mathrm{SN}}(\mathrm{Fe})\simeq 1.5\times 10^{48} \mathrm{ atoms}
\end{equation}

\begin{figure}
\plotone{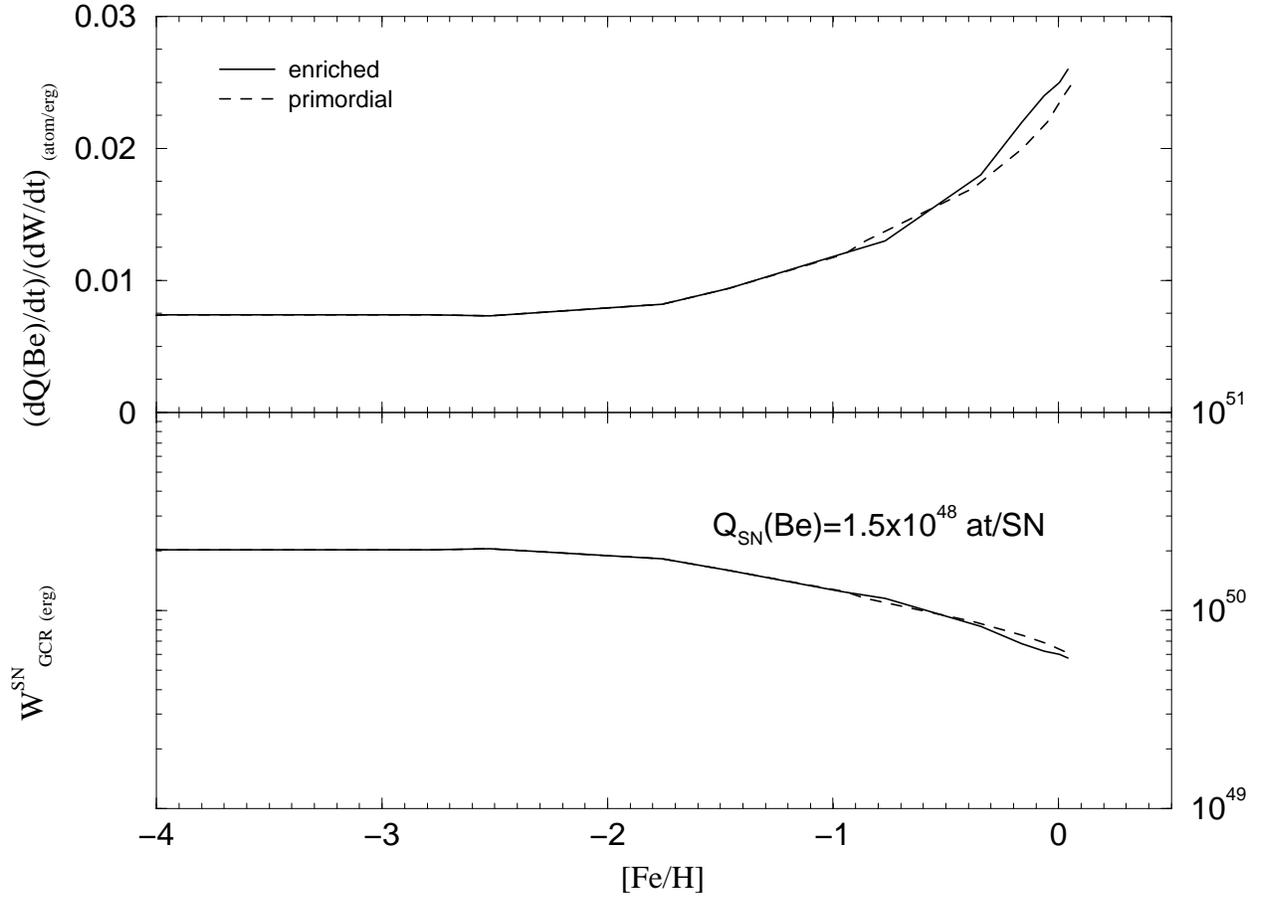}
\caption{\emph{Upper panel:} $ Q(\mathrm{Be})/\dot{W} $ vs [Fe/H]
in our $ \alpha _{\mathrm{ej}}=0.25 $ model. \emph{Lower panel:}
Energy per supernova needed to accelerate GCR, according to our evaluation of
the atoms of beryllium produced by GCR for each SN event.}
\label{energy}
\end{figure}

We have also evaluated the number of atoms of beryllium produced per erg ($\dot{
Q}(\mathrm{Be})/\dot{W}$, upper panel in Figure \ref{energy}), so we can show in the
lower panel the amount of energy per SN required to accelerate GCR, as

\begin{equation}
W^{\mathrm{SN}}_{\mathrm{GCR}}=\frac{Q_{\mathrm{SN}}(\mathrm{Be})}{\dot{Q}(\mathrm{Be})/\dot{W}}
\end{equation}

Since the early Galaxy was poor on metals, more energy per SN was required to account for the Be/Fe ratio than in the present day, when the secondary mechanism also contribute to the synthesis of Be.
In the halo phase, the value of $ W_{\mathrm{GCR}}^{\mathrm{SN}} $ is about 2$
\times  $10$ ^{50} $ erg, so the energy constraint is fulfilled because only
15--20\% of the kinetic energy of any supernova has to be involved in GCR
acceleration.

 \placefigure{energy}

\section{\label{conclusions}CONCLUSIONS}

The results of a recent and successful galactic chemical evolution model are applied
to the particular problem of the light elements. To calculate the LiBeB evolution,
it is necessary to know which are the abundances of hydrogen, helium, and the CNO
nuclei in the ISM, since they are the main targets of the GCR  to produce lithium,
beryllium, and boron. The evolution of iron is also important, because traditionally
the light element evolution has been plotted in [LiBeB] vs [Fe/H] graphs and iron
has been taken as representative of the metallicity of the stars where LiBeB have
been measured. In order to achieve this goal, we have used some of the latest and
most complete calculations of the yields from different types of stellar scenarios
for nucleosynthesis: low-- and intermediate--mass stars, Type Ia and II supernovae,
and nova outbursts.

Our calculations show that a mixture of 25\% of SN ejecta {(with upper and lower limits of 12\% and 41\%, respectively)}  and 75\% of ISM, for the
GCR composition, can account for the linear trend of BeB vs Fe. If the O/Fe
vs Fe relationship of Boesgaard et al. (1999a) (somewhat steeper than the one
obtained by ALC) is used, an acceptable fit to the Be vs O plot is also obtained.

On the other side, the lithium plateau is easily reproduced with a BBN contribution,
as usual, and the increase of the Li abundance after the halo phase is quite nicely
reproduced as well by taking into account several sources (the $ \nu  $--process,
nova outbursts, and C--stars).

Concerning the isotopic evolution, our model fails to reproduce the early--time $
^{6} $Li/$ ^{7} $Li ratio, due to a lack of $ ^{6} $Li at low metallicities. We
point out the necessity of an extra source for this isotope. The other calculated
ratios, however, nicely reproduce the data.

Finally, the energy problem has been addressed, in order to check that not too large
a fraction of the supernova energy is required to produce the light elements.

\begin{acknowledgements}
This work has been supported by the DGESIC grant PB98--1183--C03--01 and the DGI grant AYA2000--0983.
We thank {late} Dr. R. Ramaty for kindly providing us with his
numerical code for cosmic--ray induced nucleosynthesis of the light elements.
\end{acknowledgements}

\end{document}